\documentclass[manuscript]{acmart}
\AtBeginDocument{%
  }

\setcopyright{none}
\copyrightyear{2018}
\acmYear{2018}
\acmDOI{XXXXXXX.XXXXXXX}
\acmConference[Conference acronym 'XX]{Make sure to enter the correct
  conference title from your rights confirmation email}{June 03--05,
  2018}{Woodstock, NY}
\acmISBN{978-1-4503-XXXX-X/2018/06}

\usepackage{multirow}
\begin{document}
\title{Breaking the Modality Barrier: Multimodal Fusion With Missing Modalities for Cybersickness Prediction}
\author{Purnata Saha}
\email{purnata.saha@my.utsa.edu}
\affiliation{%
  \institution{University of Texas at San Antonio}
  \city{San Antonio}
  \state{Texas}
  \country{USA}
}

\author{Rifatul Islam}
\email{rislam11@kennesaw.edu}
\affiliation{%
  \institution{Kennesaw State University}
  \city{Kennesaw}
  \state{Georgia}
  \country{USA}
}

\author{John Quarles}
\email{john.quarles@utsa.edu}
\affiliation{%
  \institution{University of Texas at San Antonio}
  \city{San Antonio}
  \state{Texas}
  \country{USA}
}
\begin{abstract}
Cybersickness is a major barrier to comfortable VR use, and reliable prediction often requires multiple physiological and behavioral sensors. However, existing models typically assume a fixed set of modalities, making them vulnerable when sensors are unavailable and difficult to generalize across datasets with different sensor configurations. We adapt a transformer-based model that handles arbitrary sensor combinations without imputing missing data. Anchor-based contrastive learning maps heterogeneous sensor embeddings into a shared latent space using eye tracking as the anchor modality, while masked self-attention excludes unavailable sensors during fusion. We evaluate our approach on three publicly available cybersickness datasets: VR.net, SIM21, and Blur-Rollercoaster, collected using different VR systems and sensor configurations. Across single-sensor dropout conditions, prediction error changes by less than 0.09 MAE. Joint training across all datasets achieves 0.51 MAE and 0.63 Pearson correlation, substantially outperforming single-dataset models. These results demonstrate robust and generalizable cybersickness prediction across diverse VR systems.
\end{abstract}



\begin{teaserfigure}
  \includegraphics[width=\textwidth]{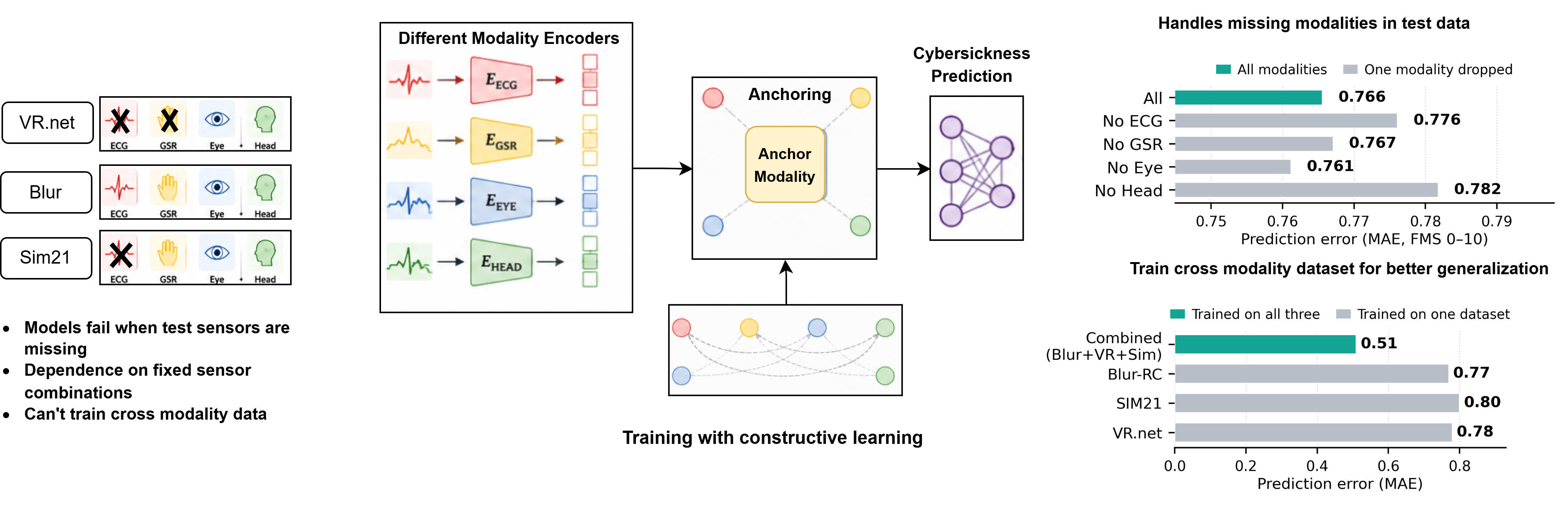}
  \caption{Architecture of the adaptive multimodal framework. Left: The training problem where heterogeneous VR datasets exhibit variable, missing modalities (ECG, GSR, Eye, Head), causing traditional pipelines to fail when sensors drop out. Center: The pipeline feeds available streams through individual temporal encoders, optimizing alignment via a contrastive anchoring layer prior to generating final cybersickness severity regressions. Right: illustrating that a single set of model weights successfully scales to handle arbitrary missing test-time sensors while showing superior generalization over single-dataset baselines and when test data have missing modalities the model can still predict based on the available one.}
  \label{fig:teaser}
\end{teaserfigure}

\maketitle

\section{Introduction}

Cybersickness prediction in virtual reality remains an open problem, largely because user responses vary across individuals \cite{lawson2021factors,biswas2024you}.  Physiological and behavioral signals such as ECG, galvanic skin response (GSR), eye tracking, and head motion have shown promise as predictive inputs \cite{jeong2023mac, islam2021cybersickness,jeong2024precyse, saha2026comparative}. Building on this, recent work has explored real-time cybersickness prediction during active VR exposure, moving toward systems capable of intervening before symptoms escalate \cite{qi2025cpnet,islam2021cybersense,hadadi2024intelligent}. However, two practical challenges limit the deployment of these systems.
First, most existing multimodal cybersickness prediction models assume that all sensor modalities used during training will also be available during inference. However, entire sensor modalities may be unavailable because a device is not present, a sensor fails, data transmission is interrupted, or users opt out of wearing certain sensors \cite{wu2024deep,shirazi2024issues}. For example, a model trained using eye tracking, head motion, GSR, and ECG may need to make predictions when only head motion and eye tracking are available. Conventional multimodal models often fail in this scenario as they are trained on a fixed set of modalities as input \cite{reza2024robust, jeong2022leveraging}. 
Second, researchers are now trying to train cybersickness prediction models using multiple datasets to improve generalization across VR applications and user populations \cite{long2025toward,jeong2024precyse, wang2025multimodal}.  However, combining datasets introduces another form of modality-level incompleteness. Different studies collect different sensor modalities and extract different feature sets, resulting in heterogeneous feature spaces \cite{setu2024mazed,hag2026predicting,ryu2026predicting}. For example, one dataset may contain eye tracking and head motion only, while another may additionally include GSR and ECG. Differences in hardware platforms, sensing technologies, sampling rates, and experimental protocols further complicate cross-dataset learning \cite{setu2026statistical}. Consequently, a model capable of utilizing all available data across datasets must be able to learn from samples with different modality combinations rather than requiring a uniform feature set. 
In this study, we utilize three recent open-source cybersickness datasets: VR.net \cite{Wen2023}, SIM21 \cite{islam2021cybersickness}, and the Blur-Rollercoaster dataset. These datasets were collected using different VR systems and sensing configurations. While all three datasets contain eye-tracking and head-motion information, the available features differ substantially. VR.net does not provide pupilometric measurements, whereas SIM21 and Blur-Rollercoaster do. The SIM21 data includes GSR signals and Heart rate at physiological data and the Blur-Rollercoaster dataset includes both GSR and ECG signals. Furthermore, the Blur-Rollercoaster dataset contains subject-level modality missingness, where complete eye-tracking recordings are unavailable for some participants due to technical failures during data collection. As a result, the combined dataset contains samples with different modality combinations, providing a realistic benchmark for studying modality-level missingness during both training and inference. 
To address these challenges, we adopt an Anchored Multimodal Physiological Deep Transformer for cybersickness prediction under missing-modality conditions inspired from previous missing modality handling architectures\cite{wu2024deep,mordacq2024adapt, jiang2025towards, wang2023multi}. The model consists of three key components. First, modality-specific encoders project physiological and behavioral signals (e.g., ECG, GSR, eye tracking, and head motion) into a shared embedding space, allowing the model to process any available subset of modalities. Second, a contrastive anchor alignment module uses a consistently available modality as an anchor and aligns representations from other modalities to improve cross-modal consistency. Third, a masked multimodal transformer fuses the available modality embeddings while handling missing inputs through attention masking rather than data imputation. The fused representation is then passed to a regression head to predict cybersickness severity.
 The main contributions of this work are as follows:

\begin{itemize}

\item We adopt a transformer-based multimodal architecture combining
modality-specific encoding and masked transformer fusion with stochastic
modality dropout. The masking mechanism accepts any subset of the modality set
at inference; in this work we evaluate it under naturally occurring
subject-level missingness during training and under single-modality dropout at
test time.

\item Using session independent cross-validation within each dataset, we conduct a modality ablation study (Scenario 1) in which the model  is trained once per dataset and evaluated under all-modality and individual modality-dropped conditions, quantifying how it performs when a specific modality is not present in test data.

\item We combine three publicly available cybersickness datasets --- VR.net, SIM21, and Blur-Rollercoaster --- collected under different VR systems and sensors. We have used  session-independent cross-validation to compare training on each dataset separately versus training on the combined dataset, to see whether combining heterogeneous datasets using this model improves generalization or not (scenario 2).

\end{itemize}

\section{Background}
In this section, we review the existing literature on missing modality handling and discuss the challenges associated with combining heterogeneous VR datasets.
\subsection{Challenges in Combining Different VR Datasets}
A big challenge in building a universal cybersickness predictor is that existing data is scattered across different independent studies \cite{bhandari2022vrise, islam2021cybersickness}. Cybersickness data is highly dependent on the individual user; previous research has shown that factors like age, gender, and prior experience with gaming or VR greatly impact how sick someone gets \cite{stanney2020virtual,kourtesis2023cybersickness,arns2005relationship}. Therefore, to make predictive models stronger and more generalizable, we need to train them on diverse datasets that expose the AI to a broad range of user profiles and VR scenarios. However, collecting such extensive data from scratch is often impractical in controlled lab settings due to time and resource constraints. Instead of trying to run a single massive study, a better solution is to combine multiple existing VR datasets collected under different environmental conditions. Yet, this is difficult because different researchers use different VR headsets, different software applications, and entirely different combinations of tracking sensors \cite{Wen2023, islam2021cybersickness}. Until now, only Setu et al. \cite{setu2026statistical} have tackled this challenge of heterogeneous data fusion by combining three distinct cybersickness datasets. To bridge the differences between these studies, they used a statistical feature abstraction method. Instead of feeding raw data into a model, they calculated simple statistical summaries, such as the mean, variance, and root-mean-square, of the sensor signals over fixed time windows. By converting mismatched raw signals into a uniform matrix of static numbers, they were able to feed data from different studies into an ordinal regression model. Their method still struggles when an entire sensor type is missing from a dataset. If one study contains ECG data and another does not, a rigid feature matrix forces researchers to drop that valuable information entirely. Rather than dropping entire datasets or throwing away valuable sensor channels to force a uniform fit, our framework is designed to utilize all available modalities across diverse sources. The model still predicts cybersickness  even if some test sensors are missing in the test set. Adapting to whatever data are present, the model ensures a stable prediction of cybersickness without requiring data imputation or structural modifications.
\subsection{Handling Missing Modalities in Deep Learning}

Multimodal learning research has explored several strategies for handling missing modalities, both during training and inference \cite{wu2024deep}. These approaches generally fall into two categories.
The first category, Recovery-based strategies operate on the premise that reconstructing absent channels allows downstream models to function normally \cite{chartsias2017multimodal, woo2023actionmae}. At the data level, Yang et al. utilized early modality composition, such as Modal-mixup, to patch missing streams by copying data from similar same-category samples \cite{yang2024modalmixup}. With deep learning, researchers transitioned to generative frameworks to synthesize raw data or latent features from surviving signals. Chartsias et al. used autoencoders to synthesize missing magnetic resonance (MR) modalities through a shared latent space \cite{chartsias2017multimodal}. Similarly, Hussein et al. introduced SensorGAN to restore missing channels directly within wearable activity recognition pipelines \cite{hussein2024sensorgan}. However, as noted by Wu et al., explicitly generating high-frequency biological data often introduces artifacts and significantly escalates computational overhead \cite{wu2024deep}. Conversely, non-recovery methods strictly utilize available inputs without data synthesis \cite{wu2024deep}. Early strategies relied on model combinations that scale poorly as sensor networks grow. To build single-model solutions, Wang et al. proposed Deep Canonical Correlation Analysis (CCA) to maximize feature association among active modalities, enabling robust cross-dataset learning \cite{wang2021acnet}. Leveraging transformers, Mordacq et al. introduced Masked Multimodal Transformers, treating missing streams as masked tokens to calculate dependencies dynamically across active sensors without data imputation \cite{mordacq2024adapt}. Extending this to physiological domains, Jiang et al. developed PhysioOmni, which utilizes a decoupled tokenizer to maintain structural compatibility with arbitrary missing inputs at inference time \cite{jiang2025towards}. By mapping sparse modalities to a consistently available baseline channel that acts as an anchor, and pairing the network with stochastic modality dropout during training, these models handle sudden sensor dropouts gracefully at runtime.



\section{Datasets}
 In this section, we will provide an overview of the three datasets
we have used in this study.
\subsection{VR.net}
The VR.net dataset \cite{Wen2023} is a large-scale, real-world data collection built specifically for Virtual Reality motion sickness research, containing a total of 165 hours of gameplay video recordings captured across 100 different VR games using the Meta Quest Pro headset. To ensure diverse visual conditions, participants in the original study were randomly assigned to play games covering multiple distinct genres. Each independent game session lasted for a fixed duration of 15 minutes, during which participants reported their FMS scores at strict 60-second intervals using a standard 1 to 5 scale to continuously assess cybersickness.

The headset recorded eye tracking data, including gaze direction, eye position, angular velocity, and eye velocity, though it is important to note that this dataset does not contain any pupilometric information. Additionally, the headset recorded the exact position, orientation, and spatial velocity of the user's head. For our specific study, we filtered and utilized 394 of these sessions to build a targeted dataset comprising exclusively the eye tracking streams, head tracking telemetry, and their corresponding time-synchronized FMS labels.
\subsection{Sim21}
SIM21 \cite{islam2021cybersickness} is a multimodal virtual reality cybersickness dataset collected during a navigation-based VR task intended to induce symptom responses under 5 immersive environments.  In the \textbf{Beach City} simulation, participants used controllers to navigate and shoot targets with a bow and arrow, choosing between teleportation or natural walking for movement. The \textbf{Road Side} simulation restricted navigation to natural walking, though the target shooting task remained similar. In the \textbf{Furniture Shop} environment, participants used teleportation to locate a hidden longbow and shoot targets. The \textbf{SeaVoyage} simulation involved seated participants navigating a boat around checkpoints using a single controller. Finally, the \textbf{Roller Coaster} simulation provided a seated experience of a roller coaster ride with no control over the movement. The dataset contains synchronized behavioral and physiological measurements, including head tracking, eye tracking, heart rate, and galvanic skin response, along with FMS from the scale 0-10. SIM21 used an HTC Vive Pro Eye headset
equipped with a Tobii eye tracker to capture high-frequency gaze
data. The original dataset tracked 27 participants across five simulation sessions each; however, data from three participants were excluded due to tracking fragmentation or early withdrawal. Consequently, the present study uses the complete, unfragmented records from the remaining 24 participants, yielding a total of 120 sessions (24 participants $\times$ 5 sessions each).
\subsection{Blur Rollercoaster}
This dataset \cite{saha2026blur} involved 38 participants who completed three counterbalanced rollercoaster conditions: no blur, dynamic blur, and static maximum blur. Each participant completed three sessions, yielding a total of 114 sessions (38$\times$3). Every session lasted 3 minutes, during which a range of physiological signals, including ECG, GSR, eye-tracking, and head-tracking, were continuously recorded. Eye and head tracking data were captured using an HP Omnicept headset, while physiological signals were recorded using Consensys Shimmer GSR and ECG units. During each rollercoaster simulation, participants provided self-reported FMS ratings on a 0--10 scale every 30 seconds, and SSQ was administered before and after each session. Prior to each rollercoaster run, participants completed a 2-minute resting phase during which baseline physiological data were recorded. All data were time-synchronized for each participant. However, eye-tracking data were missing for 14 participants due to technical issues. Because our study addresses missing modalities, we retained all 38 participants and did not exclude those with missing eye-tracking recordings.

\section{Methodology}
The methodology is divided into 4 parts: Data processing, FMS handling, Model architecture, and training setup.
The code is available at \url{https://anonymous.4open.science/r/Missing-modality-handling-for-cybersickness-8162}.
\subsection{Data Processing}

We combine three publicly available cybersickness datasets: VR.net, SIM21, and Blur-Rollercoaster, each collected with different VR hardware, sensor configurations, and sampling rates. To make these heterogeneous sources compatible with a single model, we apply a unified preprocessing pipeline consisting of four stages: modality extraction, temporal resampling, windowing, and normalization.

 For each participant, we extract up to four physiological and behavioral modalities: ECG, GSR, eye tracking (Eye), and head motion (Head). Not all datasets provide all four modalities. Blur-Rollercoaster includes ECG, GSR, Eye, and Head signals, but a subset of participants lack complete eye-tracking recordings due to sensor failures during data collection. SIM21 provides GSR, heart rate, Eye, and Head data. VR.net provides only Eye and Head signals, derived from raw head and eye pose tuples (quaternion orientation, position, velocity, and angular velocity for the head and both eyes), which are parsed and expanded into flat numeric feature vectors. For each participant, we record which modalities are available and which are missing, producing a per-participant modality availability mask that is carried through the entire pipeline.

 Source modalities are recorded at different native sampling rates. To align all modalities on a common time axis, every modality is resampled to a target frequency of 1 Hz. Modalities sampled above 1 Hz are downsampled using Fourier-based resampling using scipy. This 
was done for two reasons: primarily to reduce data complexity and
processing overhead, and to avoid resampling issues reported in previous research \cite{setu2026statistical, islam2021cybersickness, setu2024mazed}. After resampling, each modality is segmented into fixed-length temporal windows of 30 seconds (30 samples at 1 Hz) using a sliding window with a stride of 30 seconds (non-overlapping windows). Each window is assigned a single regression target, taken as the FMS label at the final timestep of the window. Each feature is normalized independently using a \texttt{StandardScaler} fitted on the training split only, and the same scaler is applied to validation and test splits to prevent data leakage. 

\subsection{Handling FMS}

A central challenge in combining these three datasets is that they report cybersickness severity on different scales. Blur-Rollercoaster and SIM21 report Fast Motion Sickness (FMS) scores on a 0 to 10 scale, while VR.net reports sickness ratings on a 1 to 5 scale. A naive linear rescaling (e.g., doubling VR.net scores to map them onto a 0 to 10 range) is psychometrically problematic: it implicitly changes the meaning of the self-reported value. For example, a VR.net score of 1 (no sickness) would become 2 on a 0 to 10 scale, which corresponds to mild sickness, introducing label noise that does not reflect the participant's actual experience.

To avoid this issue while still enabling a single shared regression target across datasets, we apply dataset-specific min-max normalization, mapping each dataset's native scale to a common $[0,1]$ interval:
\[
y_{\text{norm}} = \frac{y - y_{\min}}{y_{\max} - y_{\min}}
\]
where $(y_{\min}, y_{\max}) = (0, 10)$ for Blur-Rollercoaster and SIM21, and $(y_{\min}, y_{\max}) = (1, 5)$ for VR.net. This preserves the ordinal structure and relative position of each score within its own scale (e.g., the minimum value of each scale, denoting "no sickness," always maps to $0.0$, and the maximum value, denoting "severe sickness," always maps to $1.0$), while making the regression target directly comparable across datasets \cite{Pereira2022RescalingPain}. All values are clipped to $[0,1]$ to guard against out-of-range labels. While calculating the error metrics or correlations we have rescaled the values to its respective scale according to the datasets' FMS range.
\begin{figure}
    \centering
    \includegraphics[width=\linewidth]{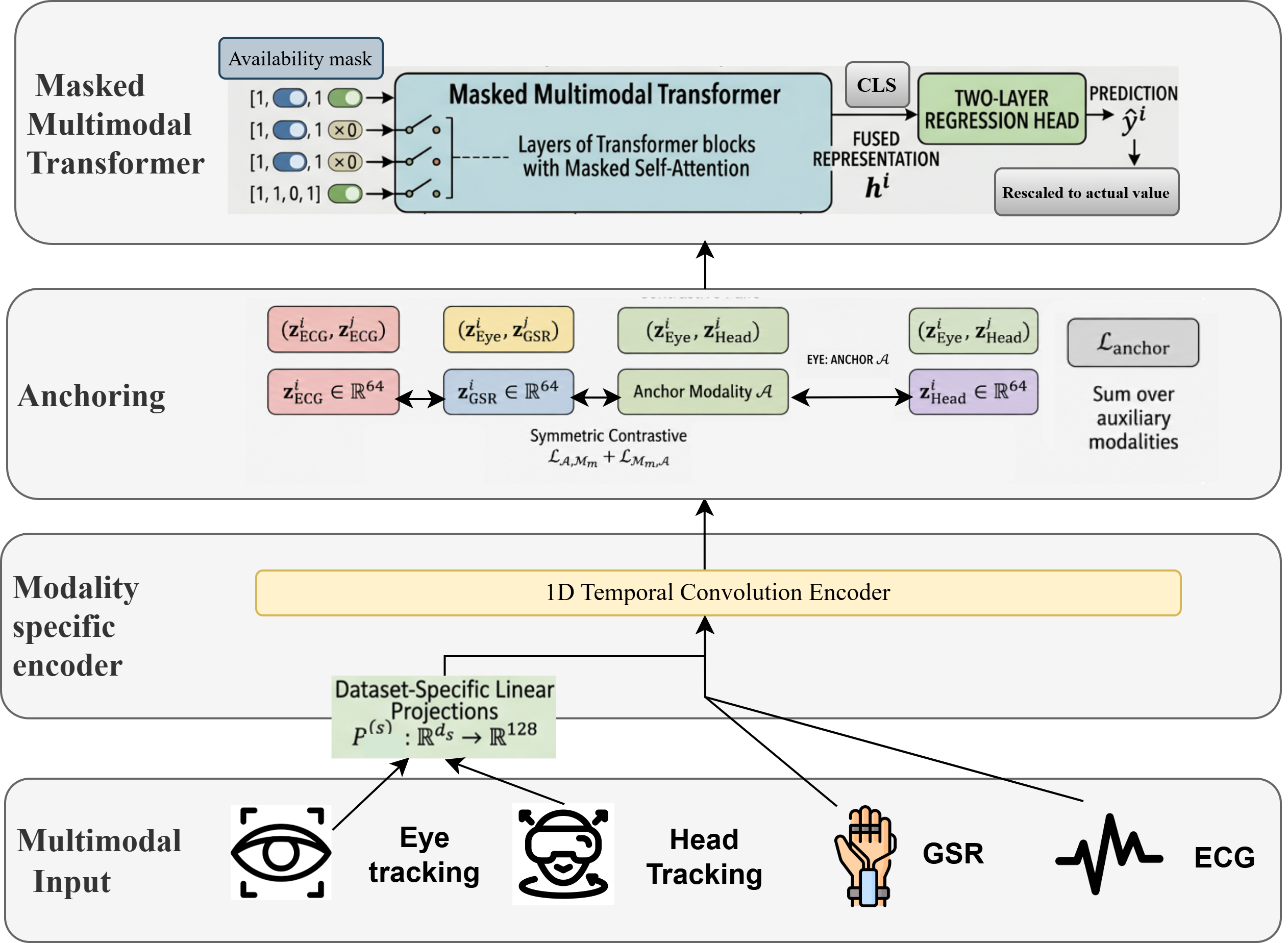}
    \caption{End-to-end framework layout. From bottom to top: Raw multimodal inputs are projected to a unified width via dataset-specific layers ($P^{(s)}$) and processed by 1D temporal convolution encoders. Disparate features are semantically mapped into a shared latent space via a symmetric contrastive anchor module ($\mathcal{L}_{\text{anchor}}$) centered on eye features. Finally, a dynamic availability mask selectively gates active sensor tokens within the transformer block to output stable cybersickness regressions}
    \label{fig:archi}
\end{figure}
\subsection{Model Architecture}

We have followed previous missing modality handling frameworks\cite{mordacq2024adapt},  and adapted it to the task of cybersickness severity regression. Let $\mathcal{D} = \{(x_m^i)_{m=1}^{M}, y^i\}_{i=1}^{N}$ denote our training set, where $M=4$ modalities (ECG, GSR, Eye, Head), $x^i = (x_m^i)_{m=1}^{M}$ is the $i$-th observation, and $y^i \in [0,1]$ is the corresponding FMS score. Given any observation $x^i$, possibly with one or more modalities missing, our goal is to train a network $F$ that predicts fms. 
The detailed architecture of the model is shown in the Figure \ref{fig:archi}.

\textbf{Modality-specific encoders.} Each modality $m$ is processed by a dedicated temporal convolutional encoder $E_m$, which takes a windowed sequence of sensor readings $x_m \in \mathbb{R}^{T \times d_m}$ (where $T$ is the number of timesteps in the window and $d_m$ is the number of features for that modality) and learns short-term temporal patterns from it.

During preprocessing, features that mean the same thing across dataset, are given the same column name, so the model treats them consistently no matter which dataset they came from. Even so, the \emph{total number} of Eye and Head features still differs between datasets, because each VR platform records a different set of measurements. VR.net provides raw motion vectors: direction, position, velocity, and angular velocity for the head and for each eye, plus raw eye-sensor readings. SIM21 and Blur-Rollercoaster instead provide pupil-based measurements: pupil diameter, pupil position, and gaze direction in several coordinate spaces. When a feature exists in one dataset but not another (e.g., pupil diameter is missing from VR.net), it is filled with zeros for the datasets that do not report it, while shared features such as gaze direction keep their actual values everywhere.

Therefore, the Eye and Head feature vectors for VR.net end up a different size than for SIM21 and Blur-Rollercoaster. To handle this in the cross dataset combination, each dataset's Eye and head features first pass through a small dataset-specific linear layer that converts them into a common size of $128$ before reaching the shared encoder. The projection layer is applied independently at each timestep, i.e., $P^{(s)}_{\text{eye}}: \mathbb{R}^{d_s} \rightarrow \mathbb{R}^{128}$ is applied to each of the $T$ timesteps of $x_m \in \mathbb{R}^{T \times d_s}$, yielding $\mathbb{R}^{T \times 128}$, thereby preserving the temporal structure for subsequent processing by the convolutional encoder. In simple terms, this layer acts as a translator: VR.net's 40-feature Eye vector and SIM21's 19-feature Eye vector are each passed through their own learned linear layer, and both come out as 128-dimensional vectors. From this point onward, the shared encoder receives inputs of the same size regardless of which dataset the sample came from, and does not need to know which dataset it originated from. The ECG signal in the Blur-Rollercoaster dataset and the GSR signals in both SIM21 and Blur-Rollercoaster contained only a single feature column. Therefore, these modalities did not require the additional translation step applied to modalities with varying feature dimensions.

\begin{figure}
    \centering
    \includegraphics[width=0.7\linewidth]{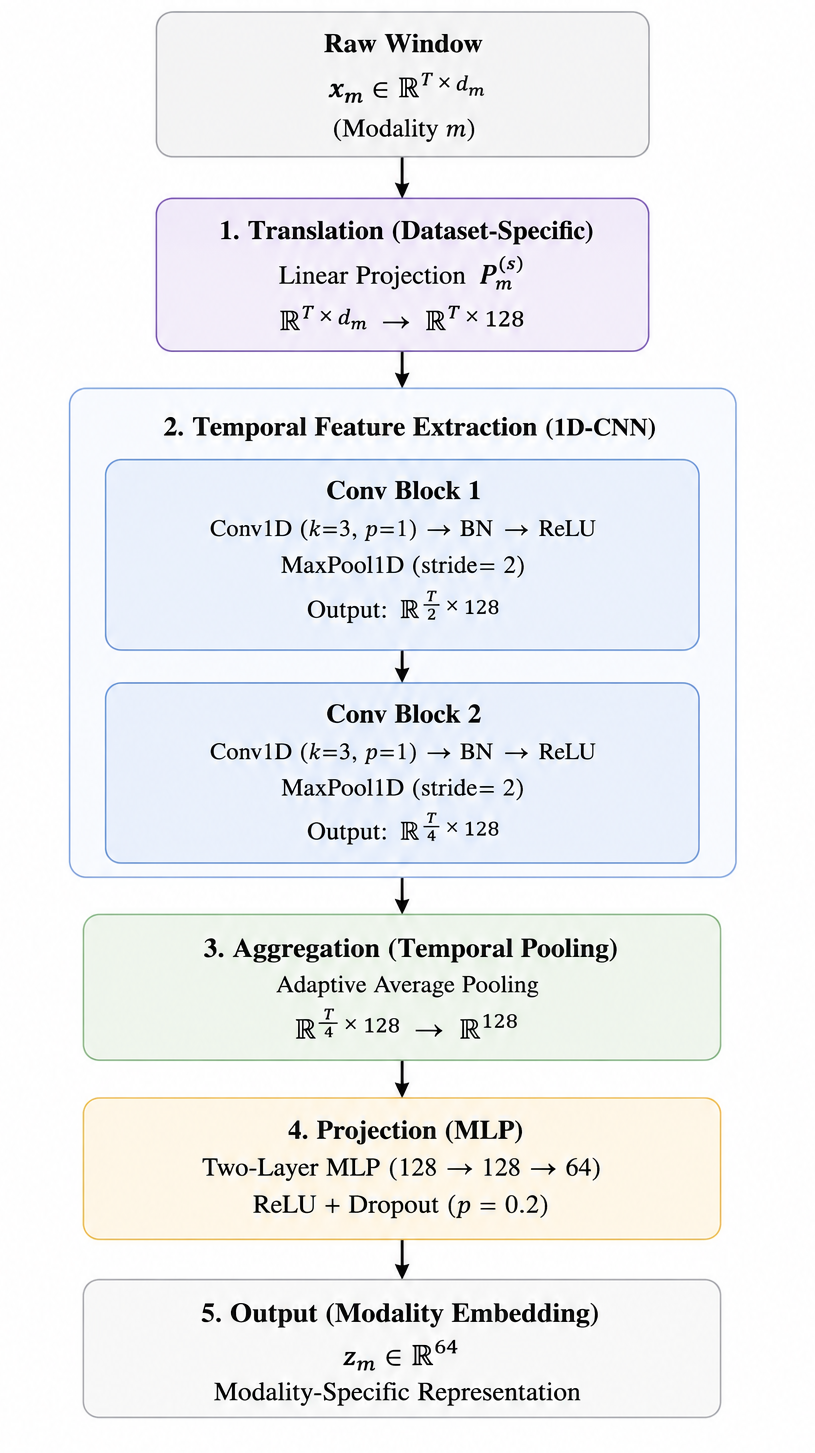}
    \caption{{Detailed architecture of the encoder used for feature extraction of the timeseries data}}
    \label{fig:encoder}
\end{figure}

After this step, the encoder applies two 1D convolutional blocks along the time axis. The detailed encoder architechture is shown in Figure \ref{fig:encoder}. Each block consists of a 1D convolution (kernel size 3, padding 1, $128$ output channels), followed by batch normalization, ReLU activation, and max pooling with stride 2, which progressively shortens the sequence while extracting local temporal patterns. An adaptive average pooling layer then compresses the remaining sequence into a single 128-dimensional vector, followed by a linear layer that keeps it at 128 dimensions. Finally, a two-layer MLP ($128 \rightarrow 128 \rightarrow 64$, with ReLU and dropout $p=0.2$) reduces this to the final modality embedding $z_m \in \mathbb{R}^{64}$, matching the shared embedding size $d=64$ used throughout the rest of the model.

\textbf{Anchoring.} We have kept Eye tracking as the anchor modality $\mathcal{A}$, since it is one of the modalities consistently available across VR.net, SIM21, and Blur-Rollercoaster. Furthermore, existing literature indicates that eye-tracking features play critical role in predicting cybersickness \cite{islam2021cybersickness, setu2024mazed}. For each auxiliary modality $\mathcal{M}_m \neq \mathcal{A}$, we align its embedding $z_m^i$ to the anchor embedding $z_a^i$ via a symmetric contrastive loss. Following Eq.~(1) of \cite{mordacq2024adapt}, for a batch of size $B$:

\begin{equation}
\mathcal{L}_{\mathcal{A},\mathcal{M}_m} = -\sum_{i=1}^{B} \log \frac{\exp(\cos(z_a^i, z_m^i)/\tau)}{\sum_{k=1}^{B} \exp(\cos(z_a^i, z_m^k)/\tau)}
\end{equation}

where $\cos(\cdot,\cdot)$ is cosine similarity and $\tau=0.07$ is a fixed temperature. We use the symmetric form $\mathcal{L}_{\mathcal{A},\mathcal{M}_m} + \mathcal{L}_{\mathcal{M}_m,\mathcal{A}}$, computed by swapping the roles of $z_a^i$ and $z_m^i$ in the denominator. The temperature $ \tau=0.07$
 controls the sharpness of the contrastive distribution; smaller values sharpen the softmax over similarity scores, producing stronger gradients that pull positive pairs (same-sample modality embeddings) closer together relative to negatives (other samples in the batch), following standard practice in contrastive representation learning \cite{mordacq2024adapt}.  Although the overall task is regression, the softmax here is used only within the contrastive alignment and does not affect the regression output. The total anchoring loss sums over all available auxiliary modalities for each sample as shown in the following equation.
\begin{equation}
\mathcal{L}_{\text{anchor}} =
\sum_{\substack{m=1 \\ \mathcal{M}_m \neq \mathcal{A}}}^{M}
\mathbb{1}[\mathcal{M}_m \text{ present}]
\cdot
\left(\mathcal{L}_{\mathcal{A},\mathcal{M}_m} + \mathcal{L}_{\mathcal{M}_m,\mathcal{A}}\right)
\end{equation}
 \textbf{Masked Multimodal Transformer.} For each sample or sequence, the four modality embeddings $z_m^i \in \mathbb{R}^d$ are stacked into a single sequence, with a learnable classification token $\text{[CLS]} \in \mathbb{R}^d$ prepended at the start. This gives a sequence $Z \in \mathbb{R}^{(M+1)\times d}$, where $M=4$. This sequence is passed through a standard transformer self-attention layer, where each token is linearly projected into a query $Q$, a key $K$, and a value $V$ (following \cite{mordacq2024adapt}), with $Q,K \in \mathbb{R}^{(M+1)\times d_k}$ and $V \in \mathbb{R}^{(M+1)\times d_v}$.

To handle missing modalities, we introduce a binary availability mask $\mathbf{a} \in \{0,1\}^{M+1}$, where $a_j=1$ if token $j$ (a modality embedding or the $\text{[CLS]}$ token) is present, and $a_j=0$ if the corresponding modality is missing for this sample. Therefore, masked (missing) modality tokens contribute zero attention weight and do not influence the output for any query. This mask is used to modify the attention computation so that missing tokens are excluded from the softmax normalization entirely, rather than simply being zeroed out afterward. Concretely, following \cite{mordacq2024adapt}, the output of a single attention head for token $i$ is:
\begin{equation}
O_i = \sum_{j} a_j \cdot \frac{\exp(Q_i^\top K_j / \sqrt{d_k})}{\sum_{\{j': a_{j'}=1\}} \exp(Q_i^\top K_{j'} / \sqrt{d_k})} \, V_j
\end{equation}

Intuitively, $O_i$ is a weighted average of the value vectors $V_j$ from all available tokens, where the weights are softmax-normalized attention scores computed only over the available tokens ($a_{j'}=1$); missing tokens are excluded from both the weighting and the normalization, so they contribute nothing to $O_i$ and do not dilute the attention given to available tokens.
 This masked self-attention is applied across  transformer layers with LayerNorm, GeLU activations, and feedforward dimension $d_{ffn}=4d=256$. The output $\text{[CLS]}$ token after the final layer, $h^i \in \mathbb{R}^d$, serves as the fused multimodal representation. The fused representation $h^i$ is passed through a two-layer feedforward regression head with ReLU and dropout, producing $\hat{y}^i \in [0,1]$.  The model is trained end-to-end to minimize:

\begin{equation}
\mathcal{L} = \mathcal{L}_{\text{MSE}} + \lambda \, \mathcal{L}_{\text{anchor}}, \qquad \mathcal{L}_{\text{MSE}} = \frac{1}{B}\sum_{i=1}^{B} (\hat{y}^i - y^i)^2
\end{equation}

where $\lambda \in \{0.1, 0.5\}$ is the contrastive weight, controlling the trade-off between regression accuracy and cross-modal alignment. We use the Adam optimizer with a learning rate of $1\times10^{-3}$, early stopping with a patience of 20 epochs on validation MSE, and a maximum of 150 epochs.

\textbf{Modality dropout.} During training, in addition to the modalities that are naturally missing for a given sample, we apply stochastic modality dropout as a data augmentation strategy. For each training sample and each modality that is actually available, we independently set $a_j=0$ with probability $p_{\text{drop}}$ and zero out the corresponding embedding $z_m^i$, regardless of whether that modality was originally observed. This means that even samples with complete sensor data are randomly presented to the model with one or more modalities artificially hidden during training.  As a result, the model is repeatedly exposed to a wide variety of missing-modality combinations during training, far more diverse than the missingness patterns that occur naturally in the dataset, which prevents the model from becoming overly reliant on any single modality and encourages it to produce stable, accurate predictions regardless of which subset of modalities is available at inference time. We set $p=0.15$ following \cite{mordacq2024adapt}.

\subsection{Training Setup}

 All models are trained using the Adam optimizer with a learning rate of $1\times10^{-3}$. The batch size is set to $8$ for Scenario 1 (per-dataset modality ablation) and to 32 for Scenario 2 which is explained in section 5. We empirically evaluated the parameter settings through a series of experiments and selected the final configuration based on validation-set performance. 
All models are implemented in PyTorch and trained on an Intel Core Ultra 9 275HX processor with an NVIDIA GeForce RTX 4090 GPU with 32 gb Ram.

\section{Experimental Result}
In this section, we present the evaluation metrics used to assess the model performance and discuss the results obtained.

\subsection{Evaluation Metrics}

Since cybersickness prediction is formulated as a regression problem, we evaluate model performance using root mean squared error (RMSE), mean absolute error (MAE), and Pearson correlation coefficient ($r$). Given the predicted cybersickness scores $\hat{y}_i$ and ground-truth scores $y_i$ for $N$ samples:
\begin{equation}
\mathrm{RMSE}=\sqrt{\frac{1}{N}\sum_{i=1}^{N}(y_i-\hat{y}_i)^2}
\end{equation}
\begin{equation}
\mathrm{MAE}=\frac{1}{N}\sum_{i=1}^{N}|y_i-\hat{y}_i|
\end{equation}
\begin{equation}
r=\frac{\sum_{i=1}^{N}(y_i-\bar{y})(\hat{y}_i-\bar{\hat{y}})}
{\sqrt{\sum_{i=1}^{N}(y_i-\bar{y})^2}\sqrt{\sum_{i=1}^{N}(\hat{y}_i-\bar{\hat{y}})^2}}
\end{equation}
Lower RMSE and MAE indicate improved prediction accuracy, while a higher Pearson correlation coefficient indicates stronger agreement between predicted and actual cybersickness trends. RMSE is reported in the same units as the FMS scale, and penalizes large errors more heavily than MAE.
\subsection{Scenario 1: Per-dataset modality ablation.} For each individual dataset, we partition the data using a 5-fold cross-validation scheme. The folds are subject-independent for the Blur-Rollercoaster and SIM21 datasets, and session-independent for the VR.net dataset. The model is trained independently on each dataset utilizing all available modalities, with stochastic modality dropout applied during training to teach the network to handle missing inputs. During evaluation, we test four distinct conditions on the same held-out data: (1) all available modalities present, and (2)--(4) with exactly one modality (Eye, GSR, or Head tracking, depending on what is available in that specific dataset) removed. For each condition, we report the MAE, RMSE, and Pearson correlation coefficient. This scenario isolates the direct contribution of each individual stream to the overall predictive performance and quantifies how gracefully the network degrades when a sensor becomes unavailable at runtime.
 Results for this scenario, reported as mean $\pm$ std across the 5 folds, are presented in Table~\ref{tab:mean_std_results}.
\begin{table}[htbp]
\centering
\caption{Cybersickness prediction performance (mean $\pm$ std) under all-modality and missing-modality conditions. The Modalities column lists the sensor streams available in each dataset; dropout conditions are reported only for modalities that dataset provides.}
\label{tab:mean_std_results}
\small
\setlength{\tabcolsep}{3pt}
\renewcommand{\arraystretch}{0.9}
\resizebox{\columnwidth}{!}{
\begin{tabular}{llcccc}
\hline
Dataset & Modalities & Scenario & MAE $\downarrow$ & RMSE $\downarrow$ & Corr. $\uparrow$ \\
\hline
\multirow{5}{*}{Blur-R}
& \multirow{5}{*}{\shortstack[l]{ECG, GSR,\\ Eye, Head}}
& All & $0.7656\pm0.0836$ & $0.9242\pm0.0304$ & $0.1360\pm0.1133$ \\
& & No ECG & $0.7761\pm0.0859$ & $0.9280\pm0.0189$ & $0.1543\pm0.1231$ \\
& & No GSR & $0.7671\pm0.0802$ & $0.9255\pm0.0247$ & $0.1250\pm0.0651$ \\
& & No Eye & $0.7612\pm0.0740$ & $0.9120\pm0.0199$ & $0.2143\pm0.0588$ \\
& & No Head & $0.7818\pm0.0732$ & $0.9334\pm0.0064$ & $0.0681\pm0.0593$ \\
\hline
\multirow{4}{*}{Sim21}
& \multirow{4}{*}{\shortstack[l]{GSR, Eye,\\ Head}}
& All & $0.8018\pm0.0127$ & $0.9327\pm0.0199$ & $0.2848\pm0.1290$ \\
& & No GSR & $0.7996\pm0.0294$ & $0.9368\pm0.0198$ & $0.1860\pm0.1060$ \\
& & No Eye & $0.8926\pm0.0765$ & $1.0120\pm0.0132$ & $0.0926\pm0.1303$ \\
& & No Head & $0.8000\pm0.0142$ & $0.9275\pm0.0236$ & $0.3186\pm0.0760$ \\
\hline
\multirow{3}{*}{VR.net}
& \multirow{3}{*}{Eye, Head}
& All & $0.7898\pm0.0086$ & $0.9456\pm0.0046$ & $0.1593\pm0.0791$ \\
& & No Eye & $0.7974\pm0.0155$ & $0.9496\pm0.0139$ & $0.1671\pm0.0552$ \\
& & No Head & $0.8233\pm0.0274$ & $0.9725\pm0.0176$ & $0.1523\pm0.0460$ \\
\hline
\end{tabular}
}
\end{table}
\subsection{Scenario 2: Heterogeneous cross-dataset combination.} 
\begin{figure}[htbp]
    \centering
    \includegraphics[width=0.9\linewidth]{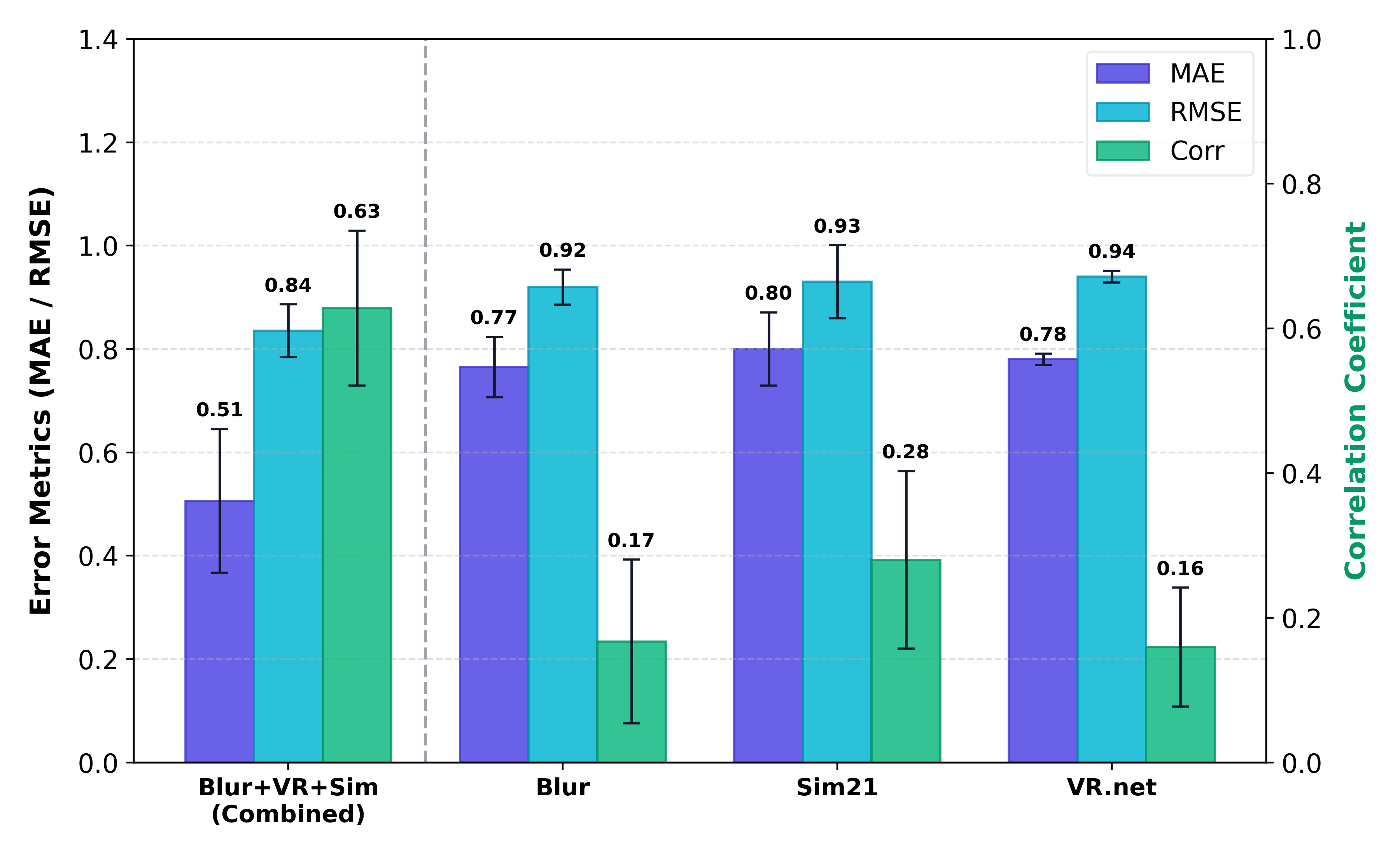}
    \caption{Overall model performance across dataset configurations. The dashed vertical line splits the joint multi-dataset model (Blur+VR+Sim) from the single-dataset models (Blur, Sim21, and VR.net). The left side tracks error metrics, while the right side tracks relationship accuracy. The mean of the 5 folds are presented in the bar graph and error bars represent standard deviation.}
    \label{fig:bar}
\end{figure}

\begin{figure}
    \centering
    \includegraphics[width=0.7\linewidth]{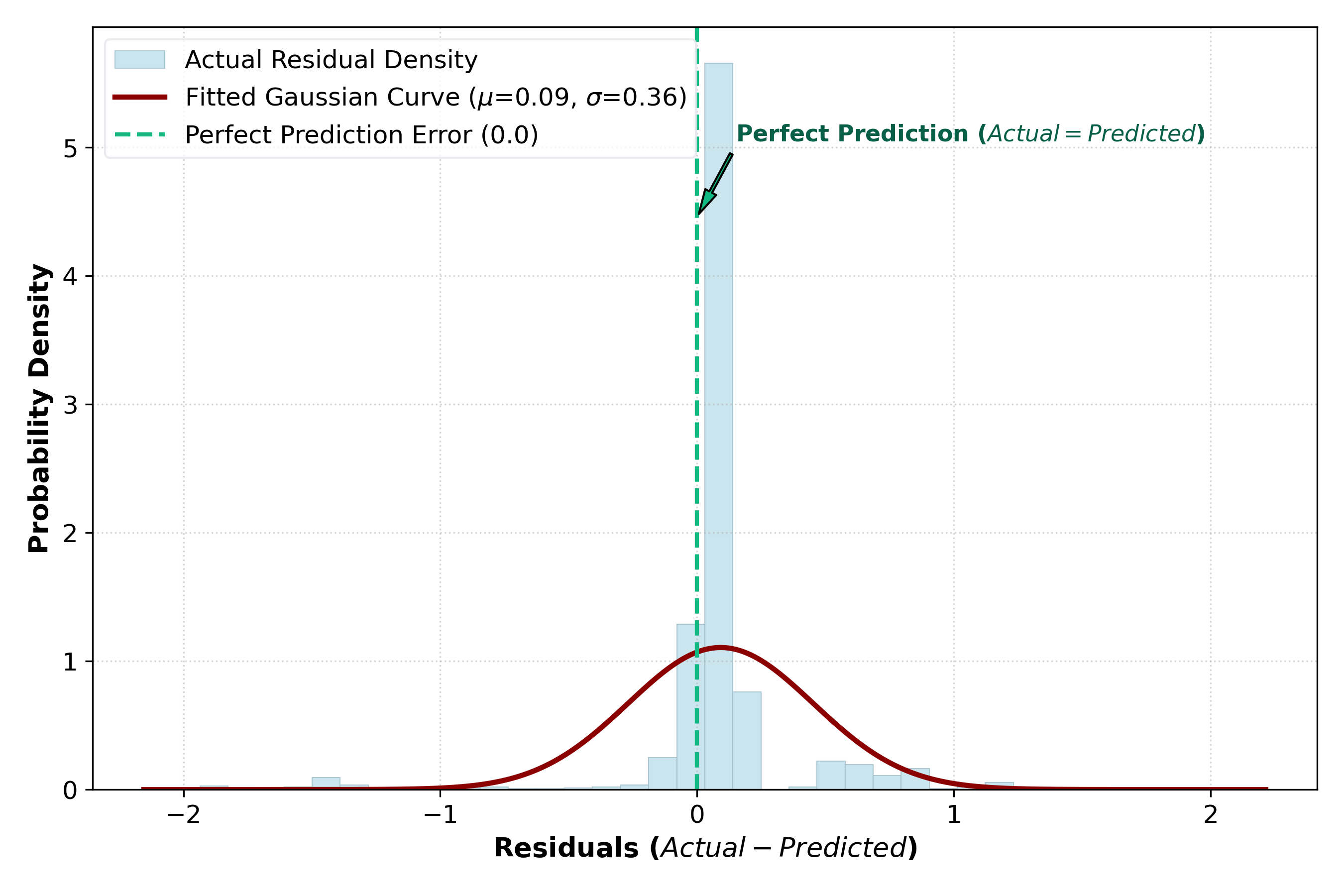}
    \caption{Residual error distribution of the cross-dataset combination model. (Residuals represent the difference between predicted and ground-truth cybersickness scores)}
    \label{fig:residuals}
\end{figure}

We compare two training conditions using subject- or session independent $k$-fold cross-validation ($k=5$). In the dataset-specific condition, the model is trained and evaluated separately on each dataset using only that dataset's participants and modalities, following the same fold structure. In the combined condition, all participants from all three datasets are pooled into a single training corpus spanning the union of all four modalities (ECG, GSR, Eye, Head); for participants from datasets lacking certain modalities (e.g., VR.net, which lacks ECG and GSR), those modality inputs are handled by the model's masking mechanism. For evaluation in the combined setting, sessions from each dataset were distributed across the folds independently, ensuring that 20\% of sessions from each dataset were reserved for testing in every fold. This session-independent $k$-fold cross-validation strategy prevents data leakage while enabling evaluation of both overall performance and dataset-specific generalization. This scenario evaluates whether pooling heterogeneous, partially-overlapping multimodal data across studies improves generalization relative to training on each dataset in isolation. Figure \ref{fig:bar} presents a performance comparison between models trained individually on each dataset versus a single unified model trained on the combined tri-dataset corpus, demonstrating robust regression even across highly discrepant sensor modalities. To further analyze prediction reliability, we examine the residual error distribution of the combined model across datasets (Figure \ref{fig:residuals}). The residual analysis illustrates the difference between predicted and ground-truth cybersickness scores and provides insight into prediction bias and error variability.

\section{Discussion}

The discussion is divided into 3 structured subsections that critically evaluate the empirical findings.

\subsection{Modality ablation}
Across all three datasets, the model showed strong stability when individual modalities were removed from the test set. For the Blur-Rollercoaster dataset, MAE remained within a narrow range across all dropout conditions. Interestingly, correlation slightly improved when Eye tracking was removed, rising from 0.14 to 0.21. We attribute this behavior to the fact that 14 of the 38 participants in the Blur-Rollercoaster dataset had missing eye-tracking recordings due to sensor failures during data collection. As a result, the all-modality condition forces the model to handle a large proportion of samples where Eye tracking is masked. When Eye tracking is explicitly dropped for all samples at test time, the model receives a more consistent input pattern across all participants, which stabilizes predictions and improves correlation. For SIM21, the most notable drop occurred when Eye tracking was removed, with MAE increasing from 0.80 to 0.89 and correlation dropping from 0.28 to 0.09. This is consistent with the literatures \cite{islam2021cybersickness,Kunduetal2025} indicating that eye-tracking features carry strong predictive signal for cybersickness. For VR.net, which only has Eye and Head modalities, removing Head tracking caused a modest increase in MAE from 0.79 to 0.82, while removing Eye tracking had a smaller effect, suggesting that head motion is a relatively stronger predictor in this dataset. Taken together, these results confirm that stochastic modality dropout during training effectively prepares the model to cope with sensor absence at test time. 

\subsection{Cross dataset combination}
A key contribution of this work is that the missing-modality-aware architecture makes it possible to combine datasets with different sensor configurations into a single training corpus, something that is not possible with conventional multimodal models that require a fixed and complete set of inputs. Basic models would need to either drop entire datasets or discard sensor channels to force a uniform input format, losing valuable data in the process. Our framework handles this naturally by masking missing modalities during training, allowing all three datasets --- VR.net, SIM21, and Blur-Rollercoaster --- to be pooled together despite having different sensor combinations. When trained on this combined corpus using session-independent cross-validation, the model becomes significantly more generalized. MAE dropped to 0.51, and Pearson correlation improved from 0.14 (mean of 3 dataset individually) up to 0.63. This improvement is likely driven by the diversity of the combined training data, which exposes the model to a much wider range of users, VR environments, hardware setups, and sickness patterns than any single dataset can provide. This matters especially for cybersickness, which is known to be highly individualized --- different people experience it differently, at different rates, and to different degrees. Training on a single dataset means the model only sees a narrow slice of this variability. By combining three datasets collected under different conditions and with different participant populations, the model is exposed to a much broader spectrum of individual sickness responses, which helps it learn more generalizable patterns rather than overfitting to the characteristics of one study.

The residual error distribution of the combined model, shown in Figure \ref{fig:residuals}, provides additional insight into prediction behavior. The residuals are approximately centered near zero with a slight positive mean of 0.09, indicating a small systematic tendency to slightly underpredict sickness severity. The standard deviation of 0.36 reflects the variability in prediction error across samples and sessions. The near-Gaussian shape of the residual distribution suggests that prediction errors are not systematically biased toward particular sickness levels or datasets, which is a positive sign for the model's generalization behavior. Some residuals fall in the tails of the distribution. These outliers are likely linked to sessions with rapid sickness onset. Alternatively, they may represent participants with unusually strong or weak responses that the model could not anticipate using only sensor data. This is a fundamental limitation of any data-driven approach to predicting a highly individualized response.

The closest prior work on cross-dataset cybersickness prediction is Setu et al.\cite{setu2026statistical}, who combined three cybersickness datasets using a statistical feature abstraction approach. Their method computed fixed statistical summaries such as mean, variance, and root-mean-square over time windows, converting heterogeneous raw signals into a uniform feature matrix that could be fed into an ordinal regression model. This approach has a fundamental limitation: if an entire modality is missing from a dataset, it must be dropped from the feature matrix entirely. For example they didn't include GSR, although they had it for two datasets. Our approach handles this differently. Rather than forcing all datasets into a rigid uniform format, we allow each dataset to contribute whatever modalities it has, and the masking mechanism handles the rest. This means no data is discarded and no sensor channel is wasted. Additionally, statistical abstraction collapses the temporal structure of the signal into a single feature vector per window, losing the fine-grained temporal patterns that our 1D convolutional encoder explicitly captures. Our framework operates on the full temporal sequence within each window, allowing it to learn dynamic within-window patterns such as gaze drift acceleration or GSR rise rate that static summary features cannot represent. The performance improvement of our combined model over individual baselines demonstrates that this richer temporal representation, combined with flexible modality handling, produces more generalizable predictions.

\subsection{Model Architecture}

The overall framework used in this study is directly inspired by the ADAPT architecture proposed by Mordacq et al. \cite{mordacq2024adapt}, which introduced the combination of modality-specific encoders, contrastive anchor alignment, and masked multimodal transformer fusion for handling missing modalities in physiological signal analysis. Our contributions on the architectural side are focused on targeted modifications rather than redesign. We replaced the original modality-specific encoders with 1D temporal convolutional encoders better suited to the time-series signals, which have different temporal dynamics and feature dimensions than the signals ADAPT was originally designed for. We introduced dataset-specific linear projection layers before the shared encoder to handle the varying feature dimensions of Eye and Head signals across VR.net, SIM21, and Blur-Rollercoaster, which was not in the original framework, since it assumed consistent feature dimensions within a single dataset. The choice of Eye tracking as the anchor is motivated by two factors. First, it is the only modality present across all three datasets, making it the most reliable reference point for alignment. Second, existing literature consistently identifies eye-tracking features as among the strongest predictors of cybersickness, typically stronger than head tracking, which was also present in all datasets.
Finally, we tuned key hyperparameters including the contrastive weight, modality dropout rate, learning rate, and early stopping patience through empirical validation on our datasets, as the original model settings were optimized for a different domain and signal type. These adaptations collectively make the framework suitable for the cybersickness prediction problem while preserving the core architectural strengths of the ADAPT model.

\section{Limitations and Future Directions}

Our study has several limitations that should be acknowledged. First, we did not validate the model in a real-time inference setting. All experiments were conducted offline on pre-recorded sessions, and it remains unknown whether the model can produce reliable predictions fast enough to be useful during an active VR session. Real-time deployment would require additional engineering and latency testing that is beyond the scope of this work. Second, the target distribution across the combined dataset is heavily imbalanced. Approximately 50\% of the time-series samples correspond to a flat zero cybersickness score, resulting in a compressed target range that limits the diversity of sickness intensities the model is exposed to during training. We did not apply any data augmentation strategy to counteract this imbalance, which may bias the model toward predicting lower or near-zero cybersickness values and limit its sensitivity to less common, higher-severity samples. Third, another key limitation of the current architecture is its resolution of missing data handling. While the proposed masked multimodal transformer effectively handles modality-level missingness---such as when an entire sensor stream is unavailable for a participant or session---it is not designed to handle sporadic or sudden missing values within an active modality's time-series window.   
Fourth, when an entire modality is absent from the training data, the model cannot handle it at test time. For example, VR.net contains no ECG or GSR data, so a model trained only on VR.net has never seen these signals and cannot make use of them even if they become available at test time. The masking mechanism only handles modalities that the model has encountered during training. This means the framework requires at least some training samples containing each modality to be able to use it at inference time.

Last, Although SIM21 provides heart rate , we did not use it because it is provided as a derived scalar value (BPM), whereas the other cardiac data in our datasets are raw ECG signals. These representations are not directly comparable and would require a separate modality-specific encoder. To keep the model architecture consistent across datasets, we therefore excluded SIM21 HR.

For future work, we suggest several directions. Validating the model in a real-time VR setting with live sensor streams would be an important next step toward practical deployment.  Finally, as more cybersickness datasets become publicly available, scaling this framework to more than three datasets would further test and potentially strengthen its generalization ability.

\section{Conclusion}
In this work we addressed two challenges in cybersickness prediction: handling missing sensors at test time and training across datasets with different sensor configurations. We adopted a transformer-based multimodal framework with contrastive anchor alignment and attention masking, and applied it to three publicly available cybersickness datasets collected with different VR hardware and sensor setups. Our modality ablation experiments showed that the model degrades gracefully when individual sensors are removed at test time, with error changing only slightly across all dropout conditions. Our cross-dataset combination experiments showed that pooling all three datasets into a single training corpus substantially improves both prediction accuracy and correlation compared to training on any single dataset alone, demonstrating that combining diverse and heterogeneous data leads to more generalizable cybersickness prediction. These results show that missing-modality-aware models are a promising path toward building cybersickness prediction systems that work reliably across different VR platforms, sensor configurations, and user populations, without requiring a fixed or complete set of sensors.



\bibliographystyle{ACM-Reference-Format}
\bibliography{sample-base}

\end{document}